\documentclass[10pt]{article}
\usepackage[OE]{express}
\usepackage{siunitx}
\usepackage{bm}
\usepackage{url}
\begin{document}
\title{Tracking capacitance of liquid crystal devices to improve polarization rotation accuracy}

\author{Rakhitha Chandrasekara,\authormark{1,*} Kadir Durak,\authormark{1} and Alexander Ling\authormark{1,2}}

\address{\authormark{1} Centre for Quantum Technologies, National University of Singapore, Block S15, 3 Science Drive 2, Singapore, 117543. \\
\authormark{2}Department of Physics, Block S12, Faculty of Science, National University of Singapore, 2 Science Drive 3, Singapore 117551.}

\email{\authormark{*}rakhitha@u.nus.edu} 



\begin{abstract}
We report a capacitance tracking method for achieving arbitrary polarization rotation from nematic liquid crystals. By locking to the unique capacitance associated with the molecular orientation, any polarization rotation can be achieved with improved accuracy over a wide temperature range. A modified relaxation oscillator circuit that can simultaneously determine the capacitance and drive the rotator is presented.
\end{abstract}



\section{Introduction}

The anisotropy of nematic liquid crystals rotates the polarization of transmitted light. The degree of rotation is a function of the crystal alignment which can be controlled by varying the amplitude of an applied AC electric field, enabling  compact electrically-variable rotators. Liquid crystal polarization rotators (LCPRs) typically rotate light in the circular polarization basis, but provide linear polarization rotation when sandwiched between correctly oriented quarter-wave plates~\cite{Chun Ye}.  

\begin{figure}[h]
\centering\includegraphics[width=5.5cm]{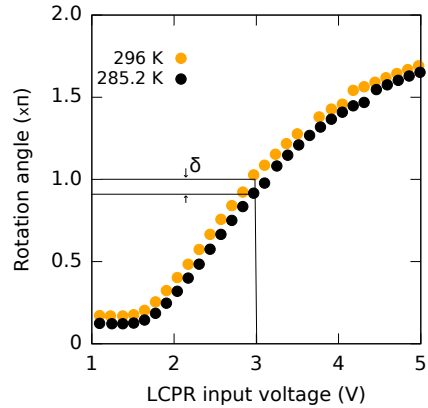}
\caption{Polarisation rotation achieved by a typical LCPR for different applied voltages at two different temperature settings. The necessary voltage to achieve a rotation of $\pi$ at \SI{296}{\kelvin} is approximately \SI{3}{\volt}, but this voltage only achieves a rotation of $0.9\pi$ at \SI{285.2}{\kelvin},  resulting in a fractional error of 10\% ($\delta = 0.1\pi$ ). As is commonly observed for most LCPRs, there is a threshold voltage (approximately at \SI{1.5}{\volt} in this case) before polarization rotation is observed.}
\label{fig:LCPR_temp_shift}
\end{figure}

A drawback of LCPRs is the temperature dependence of their dielectric permittivity. When an electric field is applied, the liquid crystal molecules respond by rotating to an angle that minimizes the free electrostatic energy \cite{Saleh}.  As the molecules have uniaxial symmetry this rotation leads to a change in the birefringence $\Delta$n resulting in polarization rotation of transmitted light. When the temperature changes the resultant change in the dielectric anisotropy $\Delta\epsilon$ affects the amount of crystal alignment needed to minimize the free electrostatic energy \cite{LC_cap_temp}. 

Therefore, if the temperature of the rotator is not taken into consideration the polarization rotation achieved under a constant applied field can be significantly different from the desired setting (Fig. \ref{fig:LCPR_temp_shift}). This can be overcome by either stabilising the LCPR's temperature \cite{LCVR}, constructing a calibration curve of the LCPR's response against temperature or implementing an optical feedback mechanism.  It is interesting to consider if improved  accuracy of a LCPR can be obtained without direct optical or temperature reference. This may enable simple and accurate operation in resource-scarce situations, e.g. on small handheld devices or onboard a small spacecraft ~\cite{tang16}.

\begin{figure}[b]
\centering\includegraphics[width=5.5cm]{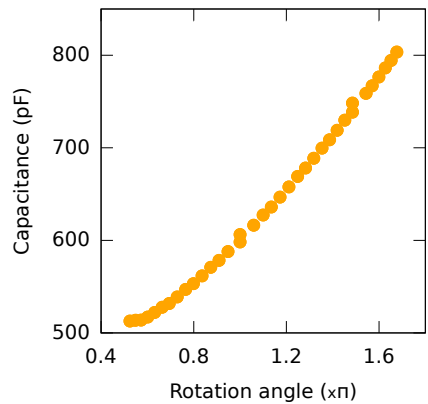}
\caption{The observed capacitance for a liquid crystal polarization rotator (LCPR) over $1.2\pi$ rad of rotation at \SI{295.4}{\kelvin}. The physical dimensions of the LCPR is measured at a clear aperture of $5\times5$\SI{}{\milli\meter} and a thickness of \SI{1.5}{\milli\meter} including the two glass plates which confine the liquid crystals. This capacitance curve is constant, within measurement error, when the temperature is between \SIrange{283}{301}{\kelvin} and can be used to achieve temperature free polarisation rotation. The temperature range above corresponds to the typical temperature experienced by our polarization instruments onboard small satellites \cite{tang16}. }
\label{fig:LCPR_cap_angle}
\end{figure}

We propose to track the capacitance of the LCPRs to improve their rotation precision even when the temperature is varying. A strong relationship exists between the LCPR's capacitance and the rotation angle which can be exploited to achieve precise polarization rotation independent of temperature. Such a relationship is shown in Fig. \ref{fig:LCPR_cap_angle} where the data was recorded at 295.4K for a device which we commonly use in our experiments. The capacitance at a rotation setting of $\pi$ rad, is approximately \SI{601}{\pico\farad} with a gradient of \SI{49.8}{\pico\farad\per\radian}. 

For most LCPRs, the capacitance curve as shown in Fig. 2 changes very slowly with temperature when operated far below the liquid crystal isotropic temperature \cite{LC_cap_temp}. The LCPRs in our study have an isotropic temperature of \SI{363}{\kelvin}. In this paper, we focus on the temperature range between \SI{283}{\kelvin} and \SI{301}{\kelvin} which represents the expected conditions for our instruments onboard small satellites \cite{tang16}. As we are operating far from the isotropic limit we can expect that the capacitance curve shown in Fig. \ref{fig:LCPR_cap_angle} to have only small changes over the working range. This is confirmed by our successful demonstration of a feedback system, over the target range, that allows LCPRs to be operated with high accuracy when locking onto the unique capacitance values.

\section{LCPR Capacitance Measurement}

\begin{figure}[h]
\centering\includegraphics[width=8.5cm]{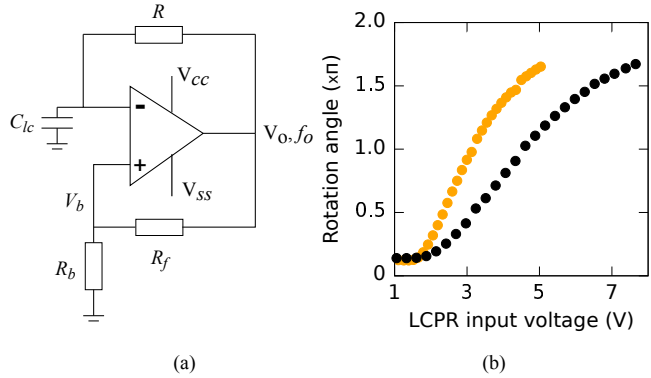}
\caption{(a) The R-C oscillator circuit with its capacitor replaced by a LCPR. The capacitance of the LCPR ($C_{lc}$) can be inferred by monitoring $f_{o}$. (b) The polarization rotation achieved by the LCPR when driven with a square wave (orange) and the R-C oscillator's charging-discharging wave (black) at \SI{295}{\kelvin}. }
\label{fig:RC_osc_response}
\end{figure}

One method of determining the capacitance of a LCPR is by using a resistor-capacitor (R-C) relaxation oscillator circuit \cite{LC_cap_temp} as shown in Fig. \ref{fig:RC_osc_response}(a). The impedance of any LCPR can be modeled by a parallel R-C circuit \cite{LC_model} whose resistance is typically in the range of few MOhms and the corresponding capacitance is 0.1-\SI{10}{\nano\farad}. Typically a LCPR is operated by applying a DC-balanced square wave (to avoid damage from ionic build-up) oscillating between 1 and \SI{10}{\kilo\hertz}. In this frequency range, the total impedance of the LCPR can be approximated by a capacitor. Thus any LCPR can be modeled as the capacitor of a R-C oscillator circuit (Fig. \ref{fig:RC_osc_response}(a)) and the output frequency ($f_{o}$) of the circuit is given by
      
\begin{equation}
  f_{o} = \frac{1}{2RC_{lc}ln(\frac{2R_{b}+R_{f}}{R_{f}})}, 
\end{equation}
where $C_{lc}$ is the LCPR's capacitance. Here $R_{b}$ and $R_{f}$ define the positive feedback voltage ($V_{b}$) present at the opamp's input terminal. This positive feedback causes the capacitor to charge and discharge towards $\pm V_{b}$ with frequency $f_{o}$. Either by setting $V_{cc} = -V_{ss}$ or using an output voltage clamp, a symmetric oscillating amplitude ($\pm V_{b}$) at the LCPR can be achieved.

\section{A New LCPR Driver Circuit}

While LCPR devices are conventionally driven by DC-balanced square waves, it should be noted that DC-balanced triangle waves also work. Both wave forms result in the same polarization rotation angle provided that the rms voltage values are the same. The voltage induced on the capacitor of the R-C relaxation oscillator is a good approximation of a triangle wave. By replacing the resistor ladder of the R-C oscillator ($R_{b}$  \& $R_{f}$) with a digital potentiometer, the amplitude of this signal and  the polarisation rotation angle can be controlled. Fig. \ref{fig:RC_osc_response}(b) shows the response of the LCPR when driven by a square wave (orange) and a triangle wave generated by a modified R-C oscillator (black) at \SI{295}{\kelvin}. The new driver is capable of achieving the same amount of polarisation rotation at higher peak to peak input voltages.

\begin{figure}[t]
\centering\includegraphics[width=8.5cm]{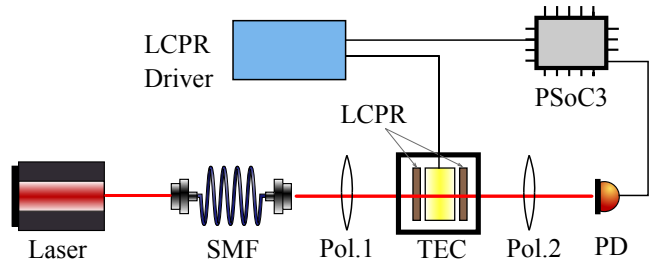}
\caption{The optical setup used to measure the achieved polarisation rotation angle with the LCPR. A collimated laser beam (at \SI{850}{\nano\meter}) coupled to a single mode fiber (SMF) is transmitted through the LCPR sandwiched between crossed polarisers (Pol.1 and Pol.2). A photodiode (PD) monitors the optical power transmitted for different rotation angles. The temperature of the LCPR is adjusted using a thermoelectric cooler (TEC). A Programmable-System-on-Chip (PSoC3) was used to drive the electronics, and to perform data acquisition.}
\label{fig:exp_setup}
\end{figure}

This modified R-C oscillator circuit was used to drive an LCPR in a polarization measurement experiment. The schematic is shown in Fig. \ref{fig:exp_setup}. The LCPR used in the test had an optimal response for light at \SI{867}{\nano\meter}. This rotator was placed between two crossed polarizers where the measured extinction ratio was 550:1. A collimated laser beam (at \SI{850}{\nano\meter}) with a photodiode (PD) have been used to measure the polarisation rotation. A thermoelectric cooler (TEC) was used to adjust the temperature of the LCPR during an experiment. 

\section{Performance Improvement and Conclusion}

\begin{table}[b]
\renewcommand{\arraystretch}{1.3}
\caption{Column 1 \& 2 show the reference capacitances and the intended rotation angles derived from Fig. \ref{fig:LCPR_cap_angle} respectively. Column 3 shows the maximum error (at \SI{283}{\kelvin}) in rotation angle without any temperature compensation. Column 4 tabulates the maximum rotation angle error when locked to the corresponding reference capacitance over the temperature range of \SIrange{283}{301}{\kelvin}. }
\label{table:table_1}
\centering
\begin{tabular}{|>{\centering\arraybackslash}m{1.5cm}|>{\centering\arraybackslash}m{2.75cm}|>{\centering\arraybackslash}m{2.75cm}|>{\centering\arraybackslash}m{2.75cm}|}
\hline
 reference cap. (pF) & intended rotation angle (rad)  & rotation angle error (rad) without any temperature compensation  &  max. rotation angle error (rad) with cap. tracking \\
\hline
601 & $\pi$ & 0.434 & 0.035 \\
\hline
633 & $1.125\pi$ & 0.360 & 0.028 \\
\hline
668 & $1.250\pi$ & 0.428 & 0.029  \\
\hline
705 & $1.375\pi$ & 0.388 &  0.034 \\
\hline
745 & $1.500\pi$ & 0.343 &  0.076 \\
\hline
\end{tabular}
\end{table}

The modified LCPR driver has the advantage of simultaneous capacitance measurement and polarisation rotation. Combining the two functions enabled precise and stable operation. A software based proportional, integrative and derivative (PID) controller has been implemented in a micro-controller (Cypress PSoC3) which can lock to the reference capacitance with a precision of $\pm$\SI{0.5}{\pico\farad}. The error signal is generated from the difference between the measured and reference capacitance values.

\begin{figure}[t]
\centering\includegraphics[width=8cm]{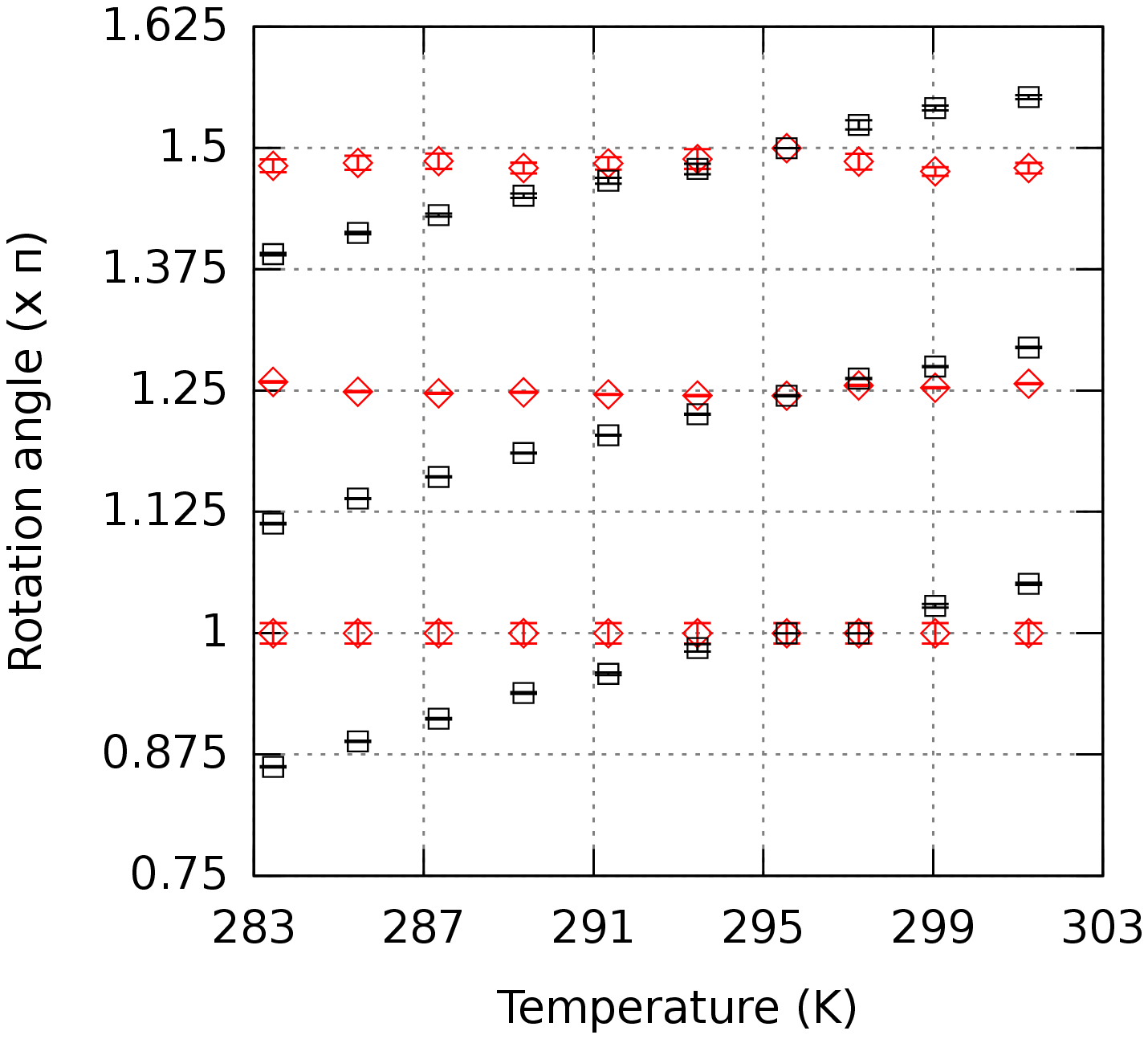}
\caption{Comparison of LCPR rotation achieved with (red data points) and without (black data points) capacitance tracking, over a range of temperature. Accurate temperature-free operation can be achieved in the range from \SIrange{283}{301}{\kelvin}. The uncertainty plotted for each locked rotation angle is obtained by propagating the $\pm$0.5 Least Significant Bit (LSB) error associated with the analogue-to-digital converter used to sample the photodiode voltage. It is interesting to observe that for intended rotation angle of 1.5$\pi$ rad the achieved rotation angle is systematically lower and this is attributed to the fast axis orientation dependence on driving voltage \cite{LC_fast_axis_volt}.}
\label{fig:rot_ang_cap_track}
\end{figure}

The performance of the LCPR with capacitance tracking enabled is presented in Fig. \ref{fig:rot_ang_cap_track} and the data was collected by the following experimental procedure. The temperature of the LCPR was set to specific values by using the TEC. Then, at each temperature setting, the PID controller was activated and checked the capacitance tracking capability for each intended angle. 

The data shows that for a temperature range of \SI{18}{\kelvin} (between \SI{283}{\kelvin} to \SI{301}{\kelvin}) where capacitance tracking enables stable polarization rotation. The performance of the LCPR with and without capacitance tracking is presented in Table 1. The maximum fractional angle error observed was 0.016 when the intended rotation angle of 1.5$\pi$ rad. In comparison without capacitance tracking the fractional angle error is 0.073 (larger by a factor of 4.5). Overall with capacitance tracking enabled the average fractional angle error was 0.007.

It is noted that the typical switching speed of the chosen LCPR is in the range of \SIrange{4}{8}{\milli\second} \cite{LC_selection_guide} and the voltage settling time of the wiper position of the chosen digital potentiometer (AD5292) is \SI{2.5}{\micro\second} \cite{AD5292_datasheet}. This enables the PID controller to settle within tens of milliseconds as the dynamic response is limited by the LCPR switching speed. Thus, the circuit can respond to temperature changes within tens of milliseconds, making the response time of the circuit sufficient for most environments, including in the small spacecraft experiments we are conducting.

The temperature performance of the system also depends on the thermal response of external components such as the charge-discharge resistor's ($R = \SI{360}{\kilo\ohm}$) temperature coefficient ($\pm$\SI{25}{ppm\per\kelvin}) and the digital potentiometer's output when operating in voltage divider mode (\SI{5}{ppm\per\kelvin})\cite{AD5292_datasheet}. These components have been carefully chosen for their low temperature coefficients to maximise the operational temperature range of the circuit. Thus, as long as the LCPR holds the capacitance curve shown in Fig. 2 the capacitance tracking method can be extended over the chosen LCPR's operational temperature range of \SIrange{0}{50}{\celsius}.

We have demonstrated a driver circuit for liquid crystal based polarization rotators which is based on a modified R-C relaxation oscillator. By tracking capacitance the rotation angle precision is improved between factor of 4.5 and 14.75 (at $1.250\pi$ rad). 
The presented R-C oscillator circuit is compact and does not require additional temperature or optical feedback. This circuit can be implemented in environments where size, weight and power are severely restricted. 

\section*{Funding}
This research is supported by the National Research Foundation (NRF), Prime Minister's Office, Singapore (Grant Number: NRF-CRP12-2013-02)

\end{document}